\documentstyle[12pt]{article}

\def\tsetrue{T}
\def\tsefalse{F}

\let\tsepaper=\tsefalse    
\let\tsenoteon=\tsefalse   
\let\tselse=\tsetrue      
\let\tseletter=\tsetrue  
\newcommand{\tsedevelop}[1]{{}} 
\newcommand{\tseprepno}{Imperial/TP/91-92/37} 
\newcommand{\tsecompldate}{16th September, 1992}

\if\tsetrue\tsepaper 
\else 
\fi
\tsedevelop{\typeout{*** T.S.E. Development mode on ***}}

\if\tsetrue\tsepaper 
\else 
\fi

\if\tsetrue\tsepaper  
\fi

\if\tsetrue\tseletter{
\typeout{          as for Letter paper ---}
}\else
\fi
\renewcommand{\theequation}{\if\tsetrue\tselse{\arabic{section}.}
\fi\arabic{equation}}
\if\tsetrue\tselse
\else
\fi

\newcommand{\vol}[1]{{\bf #1}}

\newcommand{\tselea}[1]{\tsedevelop{\nonumber \\ & { } & \mbox{ (#1) }}
\label{#1}}
\newcommand{\tseleq}[1]{\tsedevelop{\mbox{  (#1) }}\label{#1}}

\newcommand{\tbib}[1]{\bibitem{#1}\tsedevelop{ [#1] }}
\newcommand{\tref}[1]{(\ref{#1}\tsedevelop{-#1})}

\newcommand{\tcite}[1]{\cite{#1}\tsedevelop{ [#1] }}

\newcommand{\tnote}[1]{\if\tsetrue\tsenoteon \footnote{#1} \fi}
\if\tsetrue\tsenoteon{
\typeout{--- Tim Footnotes Included ---}
}\else 
\fi

\newcommand{\tcaption}[2]{\if\tsetrue\tsepaper{
\vspace{5cm} \caption{  }
}\else
{\vspace{#1} \caption{#2} } \fi }

\newcommand{\half}{\frac{1}{2}}
\newcommand{\bea}{\begin{eqnarray}}
\newcommand{\eea}{\end{eqnarray}}
\newcommand{\beq}{\begin{equation}}
\newcommand{\eeq}{\end{equation}}
\newcommand{\nnel}{\nonumber \\ {}}
\newcommand{\ennlpp}{ {} } 
\newcommand{\npagepub} {\if\tsetrue\tsepaper{\newpage }\fi}
\if\tsetrue\tsepaper{
\typeout{--- Page Breaks for Pub. Version ON ---} }\else{
\typeout{--- Page Breaks for Pub. Version OFF ---} }\fi

\newcommand{\dt}[1]{(2\pi){-3}\int d3\vec{#1}}

\newcommand{\dtk}{\dt{k}}

\newcommand{\alphabar}{\bar{\alpha}}

\newcommand{\set}[1]{\mbox{$\{ #1 \}$}}
\newcommand{\ret}[2]{R{(#1)}_{#2}}
\newcommand{\adv}[2]{A{(#1)}_{#2}}
\newcommand{\rettr}[2]{R{(#1)}_{tr,#2}}
\newcommand{\advtr}[2]{A{(#1)}_{tr,#2}}

\newcommand{\vecp}{\vec{p}}
\newcommand{\vecx}{\vec{x}}

\newcommand{\bra}[1]{\langle  #1  |}
\newcommand{\ket}[1]{| #1 \rangle}
\newcommand{\texpect}[1]{\langle \langle #1 \rangle \rangle}
\newcommand{\permalana}{\sum_{perm. \{a\} | a_N=a }}

\begin{document}

\typeout{--- Title page start ---}

\if\tsefalse\tsepaper \thispagestyle{empty}\fi

\renewcommand{\thefootnote}{\fnsymbol{footnote}}

\begin{tabbing}
\hskip 11.5 cm \= \tseprepno
\\
\> hep-ph/9209253 \\
\> \tsecompldate \\
\tsedevelop{\> (Printed out \today ) \\}
\end{tabbing}
\vskip 1cm
\begin{center}
{\Large\bf Thermal Bosonic Green Functions Near Zero Energy}
\vskip 1.2cm
{\large\bf T.S. Evans\footnote{E-mail: UMAPT85@UK.AC.IC.CC.VAXA}}\\
Blackett Laboratory, Imperial College, Prince Consort Road,\\
London SW7 2BZ  U.K.
\end{center}
\if\tsetrue\tsepaper{\begin{center}
Tel: U.K.-71-589-5111 ext. 6980 \\
Fax: U.K.-71-589-9463 \\
\mbox{  }
\end{center}
}\fi

\npagepub
\vskip 1cm
\begin{center}
{\large\bf Abstract}
\end{center}

The properties of the various types of
bosonic Green functions at finite temperature in the
zero energy limit are considered in the light of recent work.

\vskip 1cm

Nous discutons, \`a la lumi\`ere de r\'ecents travaux, des
propri\'et\'es de certains types de fonctions de Green \`a
temp\'erature finie, dans le cas limite o\`u l'\'energie
est nulle.

\vskip 1cm

\renewcommand{\thefootnote}{\arabic{footnote}}
\setcounter{footnote}{0}

\npagepub

\typeout{--- WPG1 - Main Text Start ---}
\section{Introduction}

Zero energy Green functions are used in several important physical
problems.  The most obvious is in calculations of the free energy,
the finite temperature generalisation of the effective potential.
This can be thought of as the generating functional of the 1PI
diagrams with zero four-momenta on external legs \tcite{Raybook}.
The zero-energy bosonic self-energies are also
crucial when trying to improve standard one-loop calculations of the
free energy.  More generally, the effective potential can be thought
of as the lowest order term in a derivative expansion of the
effective action \tcite{Raybook}.   Effective actions are useful
tools in their own right and to understand whether or not such a
derivative expansion can be performed at non-zero temperature means
we must know something about the analyticity of the Green functions
near zero four-momenta.

The purpose of this paper is to look at the properties of various
types of bosonic Green functions near zero energy in the light of
recent work \tcite{TSEo,TSEnpt,Ko}, in which differences between
retarded and time-ordered Green functions etc were highlighted.  This
will be done with particular  reference to a calculation of Bedaque
and Das \tcite{BD} and one in \tcite{LvW}.\tnote{and  G. and Holstein
ref. in \tcite{BD}.}

In ITF (Imaginary Time Formalism) \tcite{LvW,ITF} the energy is
discrete, $E=2\pi\imath \nu / \beta$ where $\nu$ is an integer for
bosons.   Thus one can calculate the free energy directly as the zero
energy value of bosonic Green functions is required in all methods.
The alternative approach is to calculate the Green functions at
general discrete energy and then make an analytic continuation to
complex energies (requiring boundary conditions at large energies to
be imposed on the Green functions \tcite{TSEnpt}).  One can then in
principle investigate the analyticity of the Green functions and so
the effective action near zero energy.  In zero temperature Euclidean
field theory one would obtain the same result  with this method as
when one calculates the Green functions at zero energies, i.e. the
Green functions are analytic near zero energies at zero.  One point
worth noting is that after the continuation to real energies has been
made in ITF the Green functions obtained always include the retarded
and advanced functions which are quite different at non-zero
temperature from the time-ordered functions normally encountered in
modern field theory \tcite{TSEo,TSEnpt,Ko}.

An alternative approach to thermal field theories are the RTF
(Real-Time Formalisms) \tcite{Raybook,LvW,NS,TSEnrtf}.
These involve fields at
real physical times and energies, unlike ITF.  To achieve this
one has to double the degrees of freedom so that the vertices have a
thermal label attached ($a=1,2$) and the propagators are two-by-two
matrices.  The diagrams with all external legs or vertices type one
(that is the thermal label is one) are by definition in RTF time-ordered
thermal expectation values so that the connection with standard
Minkowskii zero
temperature field theory is much closer with RTF than with ITF.

However, even in the early days of RTF, it was noticed that diagrams
can contain singular terms.  These come about because the non-zero
temperature corrections to the propagators are always proportional to
an on-shell delta function (particles in the heat bath are real not
virtual particles).  For instance the propagator for a scalar field
can be written as
\bea
\imath \Delta{ab}(k_0,\vec{k})  &=&
\imath \bar{\Delta}{ab}(k_0,\vec{k})
+ 2\pi \delta(k2-m2) n(|k_0|) D{ab}(k_0,\vec{k}) ,
\nnel
\imath \bar{\Delta}{ab}(k_0,\vec{k}) &=&
 \left( \begin{array}{cc}
\imath \bar{\Delta} & 0  \\
0 & -\imath \bar{\Delta}\ast
\end{array} \right){ab} ,
\nnel
D{ab}(k_0,\vec{k}) &=& 2\pi \delta(k2-m2) n(|k_0|)
\left( \begin{array}{cc}
1 & exp(-\beta |k_0|/2)  \\
exp(-\beta |k_0|/2) & 1
\end{array} \right){ab}  ,
\eea
where
\bea
\bar{\Delta}&=&(k2-m2 + \imath \epsilon){-1} , \; \; \;
n(z) = [exp(\beta z)-1]{-1}.
\tselea{endef}
\eea
It is possible for two or more lines in
a diagram to carry the same four-momenta and thus for singular terms of the
form $[\delta(k2-m2)]{N\geq2}$ to appear.

This can happen when a section of a diagram corresponds to a
self-energy insertion.  This leads to
contribution of the form
\beq
\sum_{c,d=1,2} n(|k_0|)2 [\delta(k2-m2)]{2} D{ac} \Sigma{cd}
D{cd}
\tseleq{esesing}
\eeq
where $D$ is the matrix coefficient of the $n \delta$ term in the
propagator and $\Sigma$ is the matrix self-energy insertion.
With
internal vertices one has to sum over all values of the thermal labels
attached to the vertices, and this is the sum over $c,d$.
Any one term in the sum in \tref{esesing}
contains singular contributions to the integrand of the form noted
above.  However when all contributions from all the different vertex
labelling are included, the singular part is cancelled leaving a
regular contribution to the Feynman diagram.  Using the matrix structure
of the propagators and self energy insertions, which follow from the
fundamental properties of thermal Green functions (KMS condition for
two-point functions), one can show that such singularities always
cancel \tcite{LvW,NS,MOU}.

The other situation where these singularities occur is when the
external leg of a diagram carries zero four-momenta.  Then we find we have a
contribution of the form
\beq
\sum_{c,d=1,2} n(|k_0|)2 [\delta(k2-m2)]{2} D{1c} R{cd}
D{1d}
\eeq
where $R$ represents the rest of the diagram.  Note that if the whole
diagram is a tadpole then $R$ is a self-energy diagram and one can use
the same proof as with general self-energy insertions to show that in
this case the singularity cancels.  For more general cases there is no
cancellation.  It is not surprising that the tadpole is a special case,
it is the only case where the external leg {\em must} carry zero four-momenta,
i.e. it is not telling us anything about analyticity near zero energy.
{}From this discussion,
it is clear that the question of the singularities in RTF is
closely related to the behaviour of Green functions near zero four-momenta.

Originally, such singularities in RTF zero four-momenta diagrams were dealt
with by the introduction of an `ad-hoc' rule \tcite{Fuj?},  namely
that one should keep only one external vertex fixed to be type one
and sum over all possible labelling of the remaining vertices,
internal and external.  Further this then gave the same result as
would be obtained when using ITF on these diagrams.  The first question
that we shall answer is this rule really `ad-hoc' and what exactly
is it doing?

With regards to the apparent singularities encountered in RTF zero
four-momenta diagrams, there is suggestion that with a suitable
regularisation there is no problem with these diagrams.
In the work of Bedaque and Das \tcite{BD} and
Landsman and van Weert \tcite{LvW}, the suggestion is that one must keep
the $\epsilon$, from the $\imath \epsilon$ terms in the propagators,
finite.  The $\epsilon\rightarrow 0$ limit is only taken at
the very end of a calculation.\tnote{LvW, below 2.4.3.}
Thus the delta functions of RTF should be replaced as
\beq
\delta(k2-m2) \rightarrow \frac{1}{\pi}
\frac{\epsilon}{(k2-m2)2 + \epsilon2 }    .
\eeq
In \tcite{BD,LvW} one-loop two-point 1PI diagrams in RTF with
both vertices fixed to be type one (so that it is a time-ordered
function) are considered.  In the case of \tcite{BD} the real part
only is investigated in a cubic self-interacting scalar theory.  The
real part of the diagram is claimed to be finite, identical to the ITF
result and analytic in both energy and three momenta  near zero
energy.  In \tcite{LvW},\tnote{LvW, 3.3.32 and environs.} the one loop
gauge boson self-energy in the Feynman gauge of a pure $SU(N)$ gauge
theory is calculated, both real and imaginary parts. Again the result
quoted for zero four-momenta is finite, pure real and identical with the
result of ITF.

The fact that the real parts are claimed to be identical with the ITF
result is not surprising in that it is an identity
that even at finite temperature, that retarded, advanced
and time-ordered functions have identical real parts for all
four-momenta (see below or
\tcite{LvW}).  Further the  $[\delta(k2-m2)]{N\geq2}$ singularity
only appears in the imaginary part so the only remarkable aspect of
the calculation in \tcite{BD} is the analyticity of the result.   The
interesting aspect of the \tcite{LvW} result is that it  is identical
to the ITF result, and in particular has a finite real part and a
zero not infinite imaginary part.   The results of these papers
suggest that if we use this $\epsilon$ regularisation, then one can
calculate the time-ordered function and it is equal to the retarded
functions (the latter is obtained using is obtained directly in ITF
or the extra or ad-hoc rule in RTF).

The results obtained using this $\epsilon$ regularisation again raise the
question, what is happening when the extra rule for such zero
energy calculations in RTF is used?  However, three further questions
come to mind.  Are the real parts of the time-ordered functions
always equal to retarded and advanced functions  at zero energy?
This is equivalent to asking are the usual results for the real parts
of diagrams in RTF and ITF at zero energy the same.  Next, are the
time-ordered or retarded functions analytic near zero four-momenta?
Lastly are the time-ordered functions, the usual RTF result, finite
at zero four-momenta and therefore calculable with a suitable
regularisation scheme?  We shall try to answer all but the last
question in the  rest of this paper.

\section{The Extra Rule In RTF}

We shall first look at exactly what is going on
with this mysterious `ad-hoc' rule for zero energy diagrams in RTF.
This `ad-hoc' rule was first derived by Matsumoto et al. \tcite{MNU}
using the Thermo Field Dynamics approach to RTF  for the case where
the Free energy is being calculated.   These authors laid the
emphasis on the fact that the free energy was {\em not} a Green
function and so one should not necessarily expect the same rules to
be used when calculating them.  Later it was shown within the path
integral approach to RTF that the generating functional, or
equivalently the partition function, for
time-independent external sources
could not be factorised in the normal manner
\tcite{TSEzm,TSEze}.  However, it was shown there that the problems
due to these additional contributions to the generating functional
could be avoided by using a simple trick.
Alternatively, one can completely avoid the need to factorise out
unwanted contributions to the partition function of RTF in the path
integral approach by using a different path in the complex time plane
\tcite{TSEnrtf}.   In either case, the derivations in
\tcite{TSEzm,TSEzm} prove that the
complete partition function in RTF can be calculated using the standard
Feynman rules {\em provided} an additional rule was used.
This is precisely the rule that had been introduced ad-hoc
elsewhere \tcite{Fuj?}.  An important point in this derivation was that
the same rule was {\em essential} for calculating the partition
function even when the external sources were static in time but
varying in space \tcite{TSEze}.  In terms of Feynman diagrams the
contributions to the partition function comes from diagrams other than
the ones with all external vertices are type one even when the
external legs carry non-zero three-momentum.  These additional
contributions are in general non-zero in such situations
though they are often zero for very simple diagrams.  None of the
diagrams are singular for general non-zero external three-momenta as
this ensures that every leg carries a unique four-momenta.

It is to be emphasised that the extra Feynman rule for zero four-momenta is
{\em not} ad-hoc.  In \tcite{MNU,TSEzm,TSEze} it was derived in very
precise situations.  In particular these papers constitute the only
derivations within RTF of how to calculate the free energy.  What is
being demonstrated by this extra rule is that there is a different
relationship between Green functions and the effective potential at
non-zero temperature from that which exists at zero temperature.

The recent work on the relationships between the various types of
real-time Green functions \tcite{TSEo,TSEnpt,Ko} throw some further
light on this extra rule.
The time-ordered expectation value of fields at real-times is
simply the RTF function with all external legs or vertices fixed to be
type one.  The $N$ retarded $N$-point Green functions, $R_a$,
are multiple commutators.  For
pure bosonic fields they are defined to be
\begin{eqnarray}
\lefteqn{R_a(t_1,t_2,...,t_N) = } \nnel &&
\permalana \prod_{j=1}{N-1} \theta(t_{a_{j+1}}-t_{a_j})
[[...[[\phi_a,\phi_{a_{N-1}}],\phi_{a_{N-2}}],...],
\phi_{a_{1}}] \ennlpp
\tselea{enptbr}
\end{eqnarray}
where $\phi_a=\phi_a(t_a,\vec{x}_a)$.  The sum is over $\{a_j\}$, $j=1$ to $N$,
running through all permutations of the numbers $1$ to $N$ subject to
the constraint $a_N=a$.
The advanced functions $A_a$ are defined in a similar manner except we
replace all the $\theta(t)$ by $\theta(-t)$ and add an overall factor
of $(-1){N-1}$, c.f. the two-point case in \tcite{LvW,ITF}.
For fields of mixed statistics see \tcite{TSEnpt}.
There is a direct link between ITF and the retarded and advanced
functions.   If the result of an ITF calculation is $\Phi(\set{z=2\pi
\nu / \beta})$ then after a suitable analytic continuation to
real energies, $\set{E}$, we find that
\bea
\PhiN(\set{E+\imath \epsilon} | \epsilon_a>0, \epsilon_{other}<0)
&=&  R_a (\{E\})
 , \nnel
\PhiN(\set{E+\imath \epsilon} | \epsilon_a<0, \epsilon_{other}>0)
&=& A_a (\{E\}),
\tselea{enitfra}
\eea
where $\set{\epsilon}$ are infinitesimal and real.

The relevant point here is that one can also relate all the RTF
functions to each of the retarded and advanced functions
\tcite{TSEnpt}.  For instance for truncated RTF functions,
$\Gamma_{tr}{\nu_1  \ldots \nu_N}$ we have
\bea
R_{tr,a}(\set{E}) &=& \sum_{\mu_j=1,2|\mu_a=1} ( \prod_{j=1}{N}
\left(
\begin{array}{cc}
1 & 0 \\
0 & e{\alphabar \beta E_j}
\end{array}
\right){\mu_j \nu_j}
\Gamma_{tr}{\nu_1 \ldots \nu_N}(\set{E})
\\
A_{tr,a} (\set{E})&=& \sum_{\mu_j=1,2|\mu_a=1} ( \prod_{j=1}{N}
\left(
\begin{array}{cc}
1 & 0 \\
0 & \sigma_j e{-\alpha \beta E_j}
\end{array}
\right){\mu_j \nu_j}
\Gamma_{tr}{\nu_1 \ldots \nu_N} (\set{E})
\eea
The parameter $\alphabar=1-\alpha$ describes how far below the
real axis the return section of the RTF curve runs.

As it is more
usual to look at 1PI functions we define
\bea
\rettr{N}{a}(\set{E}) &=&
[\ret{2}{a}(E_a) \prod_{b\neq a} \ret{2}{b}(E_b)]{-1}
\ret{N}{a}(\set{E})
\nnel
\advtr{N}{a}(\set{E}) &=&
[\ret{2}{a}(E_a) \prod_{b\neq a} \ret{2}{b}(E_b)]{-1}
\adv{N}{a}(\set{E})
\eea
where all the energies, $\set{E}$, are flowing into the diagrams,
$\ret{N}{a}$ and $\rettr{N}{a}$ are the $N$-point connected and 1PI
retarded functions respectively, with the $a$-th leg having the largest
time.  The $\ret{2}{a}$ are the retarded propagators associated with
the fields of the $a$-th leg.  Likewise for the advanced functions.
Such a definition ensures that these are the functions that obtains
when one calculates a 1PI diagram in ITF with the usual continuation
to real energies.

Now if we specialise to the case of bosonic functions and take the
zero energy limit
these expressions simplify considerably to give
\beq
R_{tr,a}(\set{E=0}) = A_{tr,a}(\set{E=0}) = \sum_{\mu_j=1,2|\mu_a=1}
\Gamma_{tr}{\mu_1 \ldots \mu_N}(\set{E
=0})  .
\tseleq{ezerartf}
\eeq
Thus it is clear that when using the extra zero-energy rule in RTF
one is calculating retarded or advanced Green
functions (they are all the same at zero energy as this proof shows)
not time-ordered functions that are normally considered in RTF, the
$\Gamma{11 \ldots 1}$.  In terms of the types of Green functions
being calculated this means that at non zero
temperature the free energy is a generating function for the
retarded and advanced functions at zero four-momenta.  This is to be
contrasted with the situation at zero temperature where the effective
potential is generating time-ordered functions.  As the non-zero
temperature proof holds at all temperatures, the zero temperature
limit tells us that the time-ordered, retarded and advanced functions
at zero four-momenta and zero energy are all identical.  In moving to non-zero
temperature, it is well known that big differences appear between the
time ordered functions and the retarded and advanced functions
\tcite{TSEo,TSEnpt}.  The only derivations of the free energy at
non-zero temperature \tcite{MNU,TSEzm,TSEzm} tell us that correct
generalisation of the zero-temperature expansion of the effective
potential in terms of real-time Green functions
is in terms of retarded and advanced functions and not
time-ordered functions.  The `ad-hoc' or `extra' rule is therefore
really neither of
these things, it is the zero-energy limit of the prescription for
obtaining retarded or advanced functions directly in RTF as
\tref{ezerartf} shows.

\npagepub

\typeout{--- WPG2 ---}

\section{Differences between Green Functions}

The methods of \tcite{TSEnpt} can be used to look at the
relation between time-ordered, retarded and advanced
thermal Green functions near zero energy.  In particular are they
equal or are the real parts equal in general?

The bosonic two-point connected
time ordered function is $\imath \Pi_{\mu \nu}{11}(t)$, where
\beq
\imath \Pi_{\mu \nu}{11}=\texpect{ T \phi_\mu(t) \phi_\nu\dagger(0)}
\eeq
and the $\mu,\nu$ indices represent any indices on the fields.  No
thermal labels are needed as the type one field are identical to the
real-time fields. The time-ordered propagator is related to the retarded and
advanced functions by
\beq
\Pi_{\mu \nu}{11}(E) = R_{\mu \nu}{(2)}(E)
+ n(E) (R_{\mu \nu}{(2)}(E) + A_{\mu \nu}{(2)}(E) ) .
\tseleq{e2pttra}
\eeq
{}From the definition of the retarded and advanced functions we have
that
\beq
\imath R_{\mu \nu}{(2)}(E) + \imath A_{\mu \nu}{(2)}(E)
= \rho_{\mu \nu}(E)
= \int dt \; e{\imath E t}
\; [ \;\texpect{\phi_\mu(t) \phi_\nu\dagger(0)} -
\texpect{\phi_\nu\dagger(0) \phi_\mu(t)  } \; ] \; .
\eeq
Using the cyclicity of the trace in the thermal expectation values
$\texpect{\ldots}$ gives the KMS relation between the thermal Wightman
functions on the right hand side.  Inserting into \tref{e2pttra}
this gives
\beq
\lim_{E \rightarrow 0}(\imath \Pi_{\mu \nu}{11}(E,\vecx) -
\imath R_{\mu \nu}{(2)}(E,\vecx))  =
\int dt \; \texpect{\phi_\mu(t,\vecx) \phi_\nu\dagger(0,\vec{0})}.
\eeq
Thus only if $\rho_{\mu \nu}(E=0)/(\beta E)$ is zero can the
time-ordered and retarded and advanced functions be equal.
{}From \tref{ezerartf} we have that
$R_{\mu \nu}{(2)}(E=0,\vec{k}) = A_{\mu \nu}{(2)}(E=0,\vec{k})$
and so $\rho_{\mu \nu}(E=0)=0$.
For instance this can be shown using spectral representation
\tcite{LvW}.\tnote{LvW, where???}
However $\rho$ may still be linear in $E$ near zero energy.

Inverting the two-point spectral representation of RTF  gives
\beq
(\Pi{-1}){11}_{\mu \nu}(E) = (1+n(E)) R{-1}_{\mu \nu}(E)
- n(E)A{-1}_{\mu \nu}(E)
\eeq
and similar manipulations as before tell us that
\beq
\lim_{E\rightarrow 0} [ (\Pi{-1}){11}_{\mu \nu}(E)
-  R{-1}_{\mu \nu}(E) ] =  - \imath
\frac{ \int dt \; \texpect{\phi_\rho(t) \phi_\eta\dagger(0)} }
{R_{\mu \rho}(0)A_{\eta \nu}(0)} .
\eeq
Looking at $\gamma(0,\vec{x})=\int dt \;
\texpect{\phi(t,\vec{x}) \phi\dagger(0,\vec{0})}$ one can
insert a complete set of energy eigenstates, $\set{\ket{n,i}}$, and find
\bea
\gamma(0,\vecp) &=& Z{-1} \sum_{n} e{-\beta E_n}
\{ \sum_{i,j} |\bra{n,i}\phi(0,\vec{0})\ket{n,j} \} |2
\geq 0 \; , \; \; \gamma(0,\vecp) \in \Re e , \nnel
Z &=& { \sum_{n} g(E_n) e{-\beta E_n} }
\eea
where we have simplified to the case of a scalar field.
The sums over $n$ run through all the distinct energy values and
$g(E_n)$ is the degeneracy of the $E_n$ energy level.   The sums over
$i$ and $j$ take the bra and ket states through all possible states
with the same energy, $E_n$ but which differ in three-momentum by $\vecp$.
At zero temperature we
can thus conclude that this thermal Wightman function is zero as
there are no vacuum to one particle processes.  However at non-zero temperature
one has a non-zero $\gamma(0)$ if there are $n$ to $n+1$ particle
processes.  In this case the heat bath is providing a background of real
physical particles which can participate in reactions precisely in
this way and thus $\gamma(0)$ is non-zero.  It is exactly this sort
involvement of arbitrary numbers of particles coming
from the heat bath in any given process that distinguishes
zero and non-zero temperature
field theory.  This is the physics that underlies the
existence of Landau damping.

Mathematically, for non-zero three-momenta, there is always a
cut across along the real axis running across the zero energy point in
the complex energy plane of non-zero temperature self-energies because
of this Landau damping.  One sees it even at the one-loop level in
simple scalar theories \tcite{We}.  While the discontinuity across
the cut is zero at zero energy $\rho(E) \rightarrow \beta E \gamma(0)$,
the Landau damping processes still ensure $\gamma$ is non-zero.
The infra-red divergence of the Bose-Einstein distribution
cancels the $E$ factor leaving the non-zero derivative of the spectral
function at zero energy, $\gamma(0,\vecp)$.  This ensures that
there is a difference between the time-ordered and retarded two-point
functions at zero energy and non-zero temperature.
Note that in the above discussion we have kept the spatial or
three-momenta dependence arbitrary.
Also note that $\gamma(0)$ is real so that $(\Pi{-1}){11}$ and
$R{-1}$ differ only in the imaginary part.

A similar analysis can be performed on bosonic three-point functions. This is
essential as the work of \tcite{TSEo,TSEnpt,Ko} shows that the
relation between retarded and time-ordered functions is much more
complicated for general functions than the special case of two-point
functions would suggest.  The pure bosonic time-ordered connected function,
$G{111}$, is related to the pure bosonic three-point retarded and advanced
functions, $\ret{3}{a},\adv{3}{a}$, through the relation
\beq
G{111}(E_1,E_2) = \sum_{cycle} n(E_2) n(E_3)
(\ret{3}{1} + e{-\beta E_1} \adv{3}{1} ) ,
\eeq
where it is implicit that the third energy variable is given by
$E_3=-E_1 -E_2$.  The sum is taken over cycles of the $1,2,3$ indices.
For general energies this suggests that the bosonic time-ordered connected
function is $O(T2)$ bigger than the bosonic retarded and advanced functions
at high temperatures.  Again we use the spectral representations for
the retarded and advanced functions.  There is a unique
function of two independent complex energy variables
$\Phi(z_1,z_2)$ given by
\bea
\Phi(z_1,z_2) &=& \dtk \sum_{cycle} \rho_1(k_1,k_2,k_3)
\frac{\imath}{z_2-k_2} \frac{\imath}{z_3-k_3} ,
\nnel
\rho_1(t_1,t_2,t_3) &=& \texpect{\phi_3(t_3)\phi_1(t_1)\phi_2(t_2)} -
\texpect{\phi_2(t_2)\phi_1(t_1)\phi_3(t_3)}
\tselea{e3ptphi}
\eea
where $z_3=-z_1-z_2$ has been defined to make the formula tidy.
The $a$-th retarded (advanced) function is obtained by letting the
$a$-th energy variable approach the real axis from above (below) and
the remaining two approach their real axes from the opposite side as
\tref{enitfra} shows.
One includes the third redundant variable $z_3$ in this scheme.  For instance
if the first field has the largest time we have the retarded function
\beq
\ret{3}{1}(E_1,E_2) = \Phi(z_1=E_1+ 2 \imath \epsilon ,
z_2 =E_2 - \imath \epsilon)
\eeq
where $\epsilon$ is a positive infinitesimal quantity.  It is
easy to show that using the cyclicity of the trace one can relate the
thermal Wightman functions in the $\rho$'s of \tref{e3ptphi}
in a generalisation of the
KMS condition on two-point functions \tcite{TSEo,TSEnpt}.  Taking the
zero energy limit as before, we look at the real part of the connected
time ordered functions $\half (G{111} + G{222})$ in the notation
used here.
At tree level in a theory with a cubic interaction
of strength $g$
between three relativistic scalar fields with dispersion relations
$\omega_a(\vecp_a)$ this would be
\beq
\half (G{111}(0,\set{\vecp}) + G{222}(0,\set{\vecp}) ) =
\frac{g}{\omega_12 \; \omega_22 \; \omega_32} .
\eeq
For the full connected three-point functions at zero energy but
arbitrary three-momenta, we find
\bea
\lefteqn{\half [G{111}(\set{0},\set{\vecp})
+ G{222}(\set{0},\set{\vecp}) ] }
\nnel
&=& \half [ \ret{3}{a}(\set{0},\set{\vecp})
+ \adv{3}{a}(\set{0},\set{\vecp}) ] +
\rho_a(\set{0},\set{\vecp}) ,
\nnel
\ret{3}{1}(\set{0},\set{\vecp}) &=&\ret{3}{2}(\set{0},\set{\vecp})
= \ret{3}{3}(\set{0},\set{\vecp})
\nnel
&=& \adv{3}{1}(\set{0},\set{\vecp})=\adv{3}{2}(\set{0},\set{\vecp})
=\adv{3}{3}(\set{0},\set{\vecp}) ,
\nnel
\rho_1(\set{0},\set{\vecp})&=&\rho_2(\set{0},\set{\vecp})
= \rho_3(\set{0},\set{\vecp}).
\eea
It is not obvious if $\rho_a(\set{0},\set{\vecp})$ should be zero when
quantum and thermal corrections are included.
Inserting a complete set of energy and three-momentum eigenstates
we see that the three-point thermal Wightman functions at zero energy
are given by
\bea
\lefteqn{\gamma_{312}(\set{0}, \vecp_1,\vecp_2) } \nnel
&=& \int d4x_1   d4x_2 d4x_3 \texpect{\phi_3(t_3,\vecx_3)
\phi_1(t_1,\vecx_1) \phi_2(t_2,\vecx_2) } \prod_{j=1}3 e{- \imath
\vecp_j . \vecx_j}
\nnel
&=& Z{-1} \sum_n e{-\beta E_n} Tr \{ V_3 V_1 V_2 \}
\\
(V_a ){ij} &=& \bra{i} \phi_{a}(0,\vec{0}) \ket{j}
\tselea{eVdef}.
\eea
Again the sum over $n$ runs through all the possible energy
values and the trace runs through all the degenerate energy states of
energy $E_n$ and where the three-momentum of the bra state differs by
$p_a$ from that of the ket state in the definition of the $V_a$ matrix
in \tref{eVdef}.  Just as in the two-point case, it is the Landau damping
phenomena present at all non-zero temperatures that ensures that the
thermal Wightman functions are not zero.

However, unlike the two-point case it is a spectral function $\rho_a$ and
not the thermal Wightman functions that appear as the difference
between the retarded and time-ordered Green functions at zero energy.
All the spectral functions are equal and can be written as
\bea
\rho_{a}(\set{E}=0, \vecp_1,\vecp_2)
&=& Z{-1} \sum_n e{-\beta E_n} Tr \{ V_3 [V_1, V_2] \}.
\eea
The trace can be rewritten as any one of the $V$ matrices multiplied
by the commutator of other two.  In general the
$V$ matrices will be non-zero, unequal and non-commuting.
If however two of the
fields are equal and the three-momenta are zero then it is an identity
that two of the matrices are equal.

Therefore at zero energy the real part of the connected three-point
functions are not generally equal to the real part of the
retarded and advanced
three-point functions at non-zero temperatures. One special case
is where two of the fields are identical and the three-momenta are
zero when it is an identity that the spectral functions are zero.
This is especially relevant as this is precisely what appears in free
energy calculations as the external legs of relevant diagrams all
correspond to the scalar field associated with the order parameter.

While the only derivations of the free energy at non-zero temperature
link it to retarded and advanced functions \tcite{NS,MNU,TSEzm,TSEze},
it may, on the evidence of
two-and three-point functions, be possible to express it in terms of
the real parts of the time-ordered functions.  It is to be stressed
that such a relationship between time ordered functions
and the free energy has yet to be proven.

\section{Analyticity}

The result of \tcite{BD} was that the one-loop bubble diagram in a
pure self-interacting cubic scalar theory was analytic about zero
four-momentum.
In considering the analyticity of Green functions near zero energy,
it is simplest to look at the $\Phi$ function
that contains all the retarded and advanced functions.  At least for
two- and three-point functions, these can then be related to the time-ordered
functions \tcite{TSEo,TSEnpt}.    What we have shown is that the zero energy
limit is very special as there the retarded and advanced functions are
all equal, the discontinuities across the cuts on the real energy axes
disappear.  However, $\Phi$ is {\it not} analytic at
$\set{E}=0$.  The discontinuities across the cuts are not in general zero at
infinitesimal real energies.  Thus there is no neighbourhood of the
zero energy point in which the function is analytic and so the
function is not analytic there.  Indeed even if the discontinuity is
exactly zero when at the zero energy point, the cuts are still running
across the zero energy point and there may well be discontinuities in
derivatives at zero energy.   All that has been shown is that the zero
energy limit of $\Phi$, with the other variables such as three-momenta held
fixed, has a unique value however it is approached, despite the fact
that the function is not analytic.  Again it is the Landau damping
processes that are causing this strange behaviour at non-zero
temperatures.  Thus the analyticity in energy reported in \tcite{BD} for the
simple bubble diagram must be incorrect as there are Landau damping
processes giving cuts across zero energy in this simple diagram
\tcite{We}.

The calculation by Weldon \tcite{We} clearly shows this cut for the
case of the one-loop scalar bubble diagram, $B$, when using ITF. The
bubble diagram is defined to be the one of the generic form
\beq
- \imath B(E,p) = \half {(-\imath g)2} (2 \pi){-4} \int d4k \;
\Delta(k;m_1) \Delta(k+p;m_2)
\tseleq{eBdef}
\eeq
with appropriate integration and propagators for whatever
formalism is being used.  Even when the masses in
the bubble diagram are equal and the three-momentum is taken to zero
as well as the energy (which is when the $[\delta(k2-m2)]2$ singularities
appear in the RTF calculation) one may or may not be sitting on an
infinitesimally short cut depending on how this limit is taken i.e.
the function does not appear to be analytic.   Two simple examples of
the real part of a bosonic scalar self-energy already exist and they
both show this lack of analyticity.  In \tcite{TSEze} the contribution
to the retarded function coming from
the scalar bubble diagram was calculated in the zero four-momentum
limit and was found to be
\bea
\lim_{E,p \rightarrow 0} B_r(E, \vecp) &=&
(T=0) + (4 \pi2){-1} \int_1{\infty} dx \; n(\beta m x)
\frac{ (x2-1)\half}{x2 + v2\gamma2}
\tselea{ebrzfm}
\\
v&=&\frac{| \vecp |}{E}, \; \; \gamma2 = (1-v2){-\half}.
\tselea{evdef}
\eea
The lack of analyticity of bosonic two-point functions can also be
seen in the leading term of the high-temperature expansion of the
one-loop self-energy of a gauge boson in any gauge theory.
This is given in many
places and is found to be of the form (e.g. see
\tcite{We2})\tnote{Eqn. (3.3)}
\bea
\Pi_t (E,p) &\propto& g2 T2 [ v{-2} +
\frac{v2-1}{2 v3} \log \left( \frac{v+1}{v-1} \right) ] +
\ldots
\tselea{etgbse}
\\
\Pi_l (E,p) &\propto& g2 T2 (1-v{-2})
[1-\frac{1}{2 v} \log \left( \frac{v+1}{v-1}  \right) ]
+ \ldots
\tselea{elgbse}
\eea
The factor of proportionality depends on the details of the gauge theory.
These results are valid when all parameters are smaller than the
temperatures so that $E,p$ need not be infinitesimal, they could be
comparable with the zero temperature masses of any particles in the theory.

In both cases these one-loop self-energies are clearly not analytic at
zero four-momentum.  No problems with any sort of singularities are
encountered in the calculations
and the real parts are identical to the time-ordered
functions, the $11$ diagrams of RTF.  Thus the real parts of $11$
RTF diagrams are not analytic
at zero four-momentum, in contradiction with the results of \tcite{BD}.

\section{Conclusions}

In this paper three questions were studied.  The `ad-hoc' rule
for zero-energy diagrams in RTF was shown to be nothing more than the
rule needed at zero energy to calculate retarded and advanced
functions.  It was also noted that the partition function is related
to an expansion of retarded and
advanced functions and {\em not} time-ordered Green functions with
which one is familiar in zero-temperature field theory.

The difference between Green functions were then studied without any
approximation being made.  For the
two-point functions it was shown that the time-ordered and  retarded
functions differ in their imaginary part by a non-zero quantity.
This contradicts the results for the one-loop QCD self-energy given in
\tcite{LvW}.  In the case of the real part of connected Green
functions a difference was found between time-ordered and
retarded functions. Generally it is not zero but in the case where two
of the fields are identical, such as is relevant to free energy calculations
the difference appears to be zero.  There is therefore a possibility
that the real parts of time-ordered and retarded bosonic Green functions
may be equal in special cases.  In particular one might be able to use
the time-ordered Green functions to calculate free-energies directly
but this has yet to be shown.

Finally, it was noted that one-loop self-energy diagrams sometimes show
explicitly a lack of analyticity at zero four-momentum.  This is
understandable on physical grounds and one will always expect to find
it at some order in the calculation.  This calls into question
derivative expansions of effective actions at finite temperature.

Overall it would seem that the $\epsilon$ regularisation stressed in
\tcite{LvW,BD} will not help.  The `problems' of analyticity and
equality of the different types of thermal bosonic Green functions at
zero energy  should be there and reflect genuine new physics not
present in the more familiar zero temperature field theory.

\vspace{1cm}
Since this work was completed, the $\epsilon$ regularisation scheme
has been studied by Weldon \tcite{We3}
and shown to be more intricate than suggested in \tcite{BD,LvW}.

\npagepub
\section*{Acknowledgements}

I would like to thank R. Kobes, G. Kunstatter and the Institute for
Theoretical Physics at the University of Winnipeg for their
hospitality.  Illuminating discussions with M.A. van Eijck, R.L. Kobes,
J.C. Taylor and H.A. Weldon are gratefully acknowledged.  I would
like to thank the Central Research Fund of the University of London
for financial support that made my visit to Winnipeg possible

\npagepub

\typeout{--- references ---}

\end{document}